\documentclass{article}
\usepackage{color,graphicx,rotating}
\usepackage[super,nomove]{cite}

\newcommand{\kms}{km~s$^{-1}$}
\newcommand{\Msun}{M$_\odot$}
\newcommand{\gtrsim}{\stackrel{\textstyle >}{\sim}}

\newcommand{\Rsun}{R$_{\rm \odot}$}

\makeatletter
\renewcommand\@biblabel[1]{(#1)}
\makeatother
\begin{document}

\title{\large THE PUZZLING CASE OF 56 PEGASI: \\
A FAST ROTATOR SEEN NEARLY FACE-ON?}
\author{\it By A. Frankowski and A. Jorissen\\
\it Institut d'Astronomie et d'Astrophysique\\
\it Universit\'e Libre de Bruxelles}
\date{}
\maketitle

{\large A spectroscopic orbit has been recently found by R. Griffin for the
long known barium star 56~Peg, which is also a strong X-ray source. This
short-period, low mass-function orbit raises several questions regarding the
history of the system.
In the present paper, we show that it is not easy to find an evolutionary
history for 56~Peg which is consistent with the
current component's masses, unless one assumes that the giant is
a relatively fast rotator (a~few times 10~\kms). The hypothesis of
fast rotation
allows us to explain some other peculiarities of this object as well.
}

\begin{flushright}
{\it [...] when you have excluded the impossible, \\
whatever remains, however improbable, \\
must be the truth.} \\
--- Sir Arthur Conan Doyle, \\
{\it The Adventure of the Beryl Coronet}
\end{flushright}

\section*{\it\normalsize Introduction}
\label{Sect:intro}
\begin{figure}
\resizebox{\hsize}{!}{\includegraphics{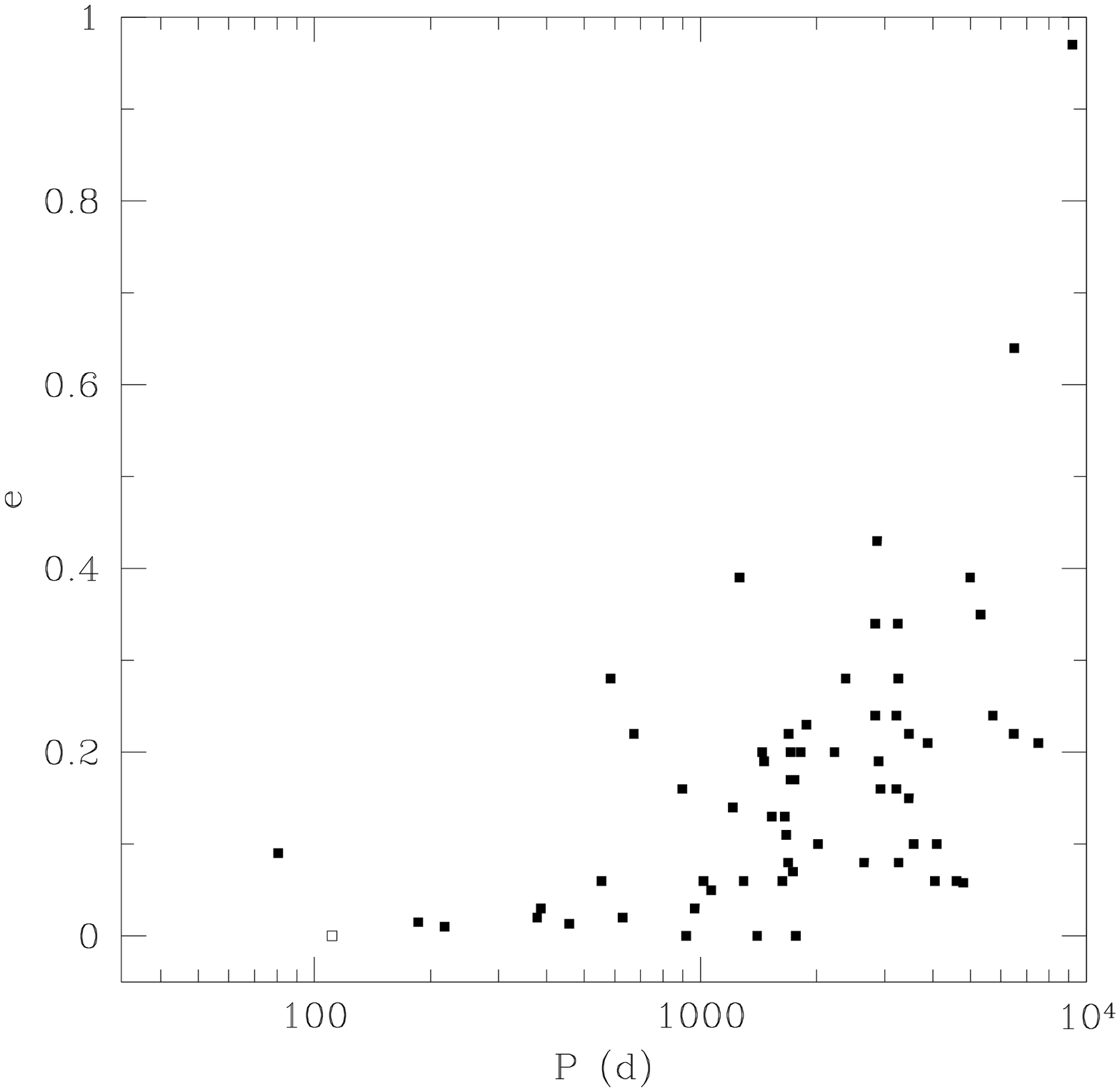}}
\caption[]{
The eccentricity -- period diagram of barium stars, with 56~Peg represented as
an open square. Data for the other stars are from Jorissen et al.
\protect\cite{Jorissen-1998}.
\label{Fig:elogP}
}
\end{figure}

\mbox{}\indent In paper 186 of R. Griffin's series on {\it Spectroscopic
Binary Orbits from Photoelectric Radial Velocities} appearing in this
issue of the {\it Magazine}\cite{Griffin-2006}, a 111.15~d orbit is
found for 56 Peg, a
rather luminous barium star which is as well a strong X-ray source.
The star 56~Peg is also remarkable in having, despite its large radius 
(the object is classified as a bright giant, listed as K0.5 II by Keenan
\& McNeil\cite{Keenan-McNeil-1989};
an interferometric measurement by van Belle et al. \cite{vanBelle-1999}
yields a radius of $40\pm4.7$~\Rsun\ for a uniform-disc approximation),
the second shortest orbital period
among barium stars, as revealed by Fig.~\ref{Fig:elogP}. This figure
displays the eccentricity -- period diagram of barium stars.
Its mass function of $3.73 \times 10^{-5}$~\Msun\ is the
smallest among barium stars. The giant is of solar or slightly subsolar
metallicity \cite{Fernandez-Villacanas-1990,Zacs-1994}: 
[Fe/H]$=-0.15\pm0.28$.
Older metallicity determinations \cite{Williams-1975,Luck-1977}, in the
range [Fe/H] $=-0.21$ to $-0.01$,  
are thus consistent with these recent values. For a thorough discussion of
the object's spectral type and abundances the reader is referred to the
review by Griffin\cite{Griffin-2006}.

An absolute magnitude of  $M_V \sim -1^m.3 \pm0.1$
is quoted by Griffin for 56~Peg, based on literature values of the $V$
magnitude and on its Hipparcos parallax\cite{ESA-1997} of $6.07
\pm0.67$~milliarcsecond\footnote{abbreviated `mas' in the remainder of this
paper} ($d=165 \;
\pm18$~pc). Distance estimates based on the straight inversion of the parallax, as done
here, are subject to statistical biases. In this specific case however, maximum likelihood
estimates of the distance (based as well on the Hipparcos  parallax, but providing unbiased
distance estimates) yield distances consistent with the above value: Mennessier et 
al. \cite{Mennessier-1997} obtain $d=161 \pm16$~pc while Famaey et al. \cite{Famaey-2005} give
$d=167~\pm18$~pc. 

In what follows, wherever clarity requires to differentiate the giant
component in 56~Peg from the system as a whole, we will use the name
56~Peg~A for the giant.

\section*{\it\normalsize Location of 56~Peg~A in the Hertzsprung-Russell diagram}
\label{Sect:HR}

\mbox{}\indent Several arguments presented in this note will be based on the
location of 56~Peg~A in the Hertzsprung-Russell (H-R) diagram. Therefore,
we prefer to rely on consistent sets of magnitude and colour measurements
from Tycho-2 \cite{Hog-2000} and Hipparcos \cite{ESA-1997}. There are 193
$V_T$ and $B_T$ measurements  available for 56~Peg, yielding $<V_T>\;
= 4.906 \pm 0.022$ and  $<B_T-V_T>\; = 1.53
\pm 0.04$, with the quoted uncertainty corresponding to the $\sigma$ of gaussians fitted to the
$V_T$ and
$(B_T-V_T)$ distributions (Hence, the standard error on the quoted mean is in fact  smaller by
the factor
$193^{1/2} = 13.9$). The average uncertainty on single $V_T$ and $B_T$  measurements of 56~Peg is
0.021 and 0.038~mag, respectively, so that there is no indication from the Tycho-2 photometry
that the star is variable (since the standard dispersions of both the $V_T$ and $(B_T-V_T)$
distributions simply reflect the measurement errors). Using Eqs.~1.3.29 and 1.3.33 of Volume 1 of
the Hipparcos catalogue \cite{ESA-1997} to convert the Tycho-2 photometry to the Johnson system, one
finally obtains $V = 4.76 \pm 0.02$ and $B-V = 1.28 \pm 0.03$.
An analogous procedure using even more accurate measurements from Hipparcos
{\it Hp} band leads to the same average value of $V$, but Hipparcos data
do reveal small intrinsic photometric variability with a scatter of
0.011~mag (Hipparcos catalogue field H46; see also the discussion in 
Griffin's paper\cite{Griffin-2006}), which is below Tycho-2 detection
limit and is also sufficiently small not to pose a threat to our further
reasoning.
The obtained value for the $V$ magnitude
falls in the literature range $V = 4.7$ -- 4.8 quoted by Griffin, and yields
$M_V = -1.32 {+0.23 \atop -0.26}$ (where the quoted uncertainty combines the
uncertainty on the parallax and on the photometry, but is in practice
completely dominated by the parallax formal error).

\begin{figure}
\resizebox{\hsize}{!}{\includegraphics{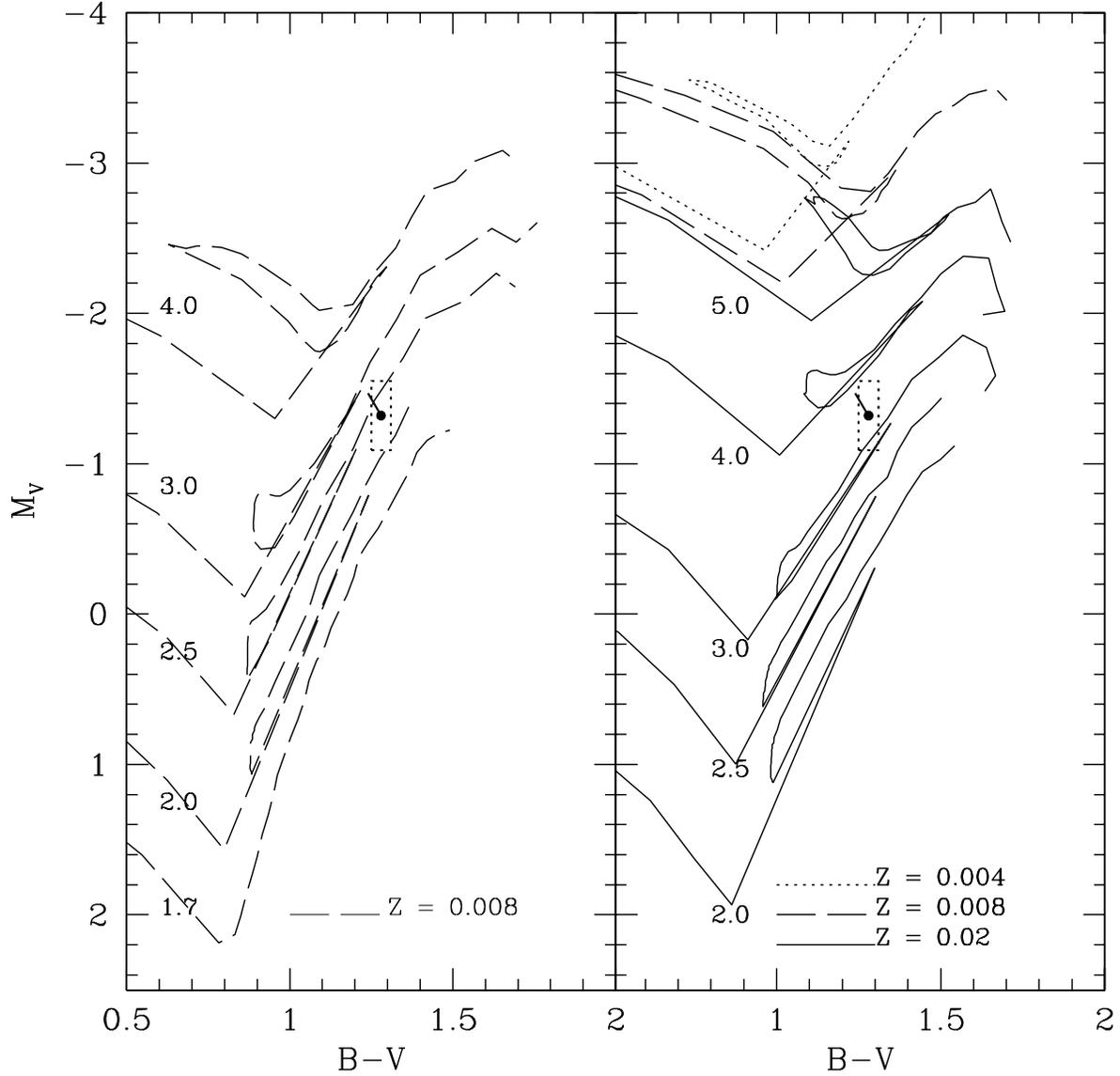}}
\caption[]{
Position of 56~Peg~A from Hipparcos and
Tycho-2 photometry (black dot within an error box; note that the formal
error in $M_V$ is completely dominated by the parallax uncertainity)
in the observational H-R diagram, plotted against Geneva theoretical
tracks \protect\cite{Lejeune-Schaerer-2001}, of various metallicities as
indicated
in the figure. Labels indicate initial masses of the stars. The thick oblique
line protruding from the  observed location of 56~Peg~A
corresponds to the de-reddening vector for $A_V = 0.15$~mag, as derived from
the extinction model of Arenou et al. \protect\cite{Arenou-1992}.  
}
\label{HRtracks.fig}
\end{figure}

As the companion is much fainter that the giant and is not seen in the
visible light (though it has been detected in the UV\cite{Griffin-2006}),
we note that it is safe to ascribe the above photometry to the giant star.
These values have been used in Fig.~\ref{HRtracks.fig} to locate 56~Peg~A
in the observational H-R diagram displaying stellar evolution tracks from
Lejeune \& Schaerer \cite{Lejeune-Schaerer-2001}, computed up to the first
thermal pulse on the asymptotic giant branch (AGB).
They reveal that 
in the solar-metallicity case ($Z = 0.02$), the giant star in the 56~Peg
system has a mass of the order of 3--4~\Msun, whereas the mass may be as
low as 2~\Msun\ for the subsolar metallicity of $Z= 0.008$ (corresponding
to [Fe/H]~$=-0.4$). 
Admittedly, such estimates may depend upon the details of the
complex theoretical modelling of late stages of stellar evolution.

More importantly, as we will now show, 56~Peg~A has not attained the
thermally-pulsing asymptotic giant branch (TP-AGB) yet. Therefore, the  nucleosynthesis associated
with thermal pulses and the ensuing third dredge-ups cannot be invoked to account for the barium
enrichment observed in the atmosphere of 56~Peg~A. 

From the observed parameters of 56~Peg~A ($B-V = 1.3$ and $M_V = -1.3$)
applying a bolometric correction one gets a luminosity of
$\log L/L_{\odot} = 2.7$.
The luminosity at the onset of the TP-AGB has been estimated in the
following way. In the solar-metallicity case 
($Z = 0.02$; right panel of Fig.~\ref{HRtracks.fig}), the core-mass --
luminosity relationship from Bl\"ocker \cite{Blocker-1993} (as given by
Eq.~1 of Herwig et al. \cite{Herwig-1998}) predicts $\log L = 3.5$ at the
start of the TP-AGB for a 3~\Msun\ star, when the core mass amounts to
0.54~\Msun\ according to Fig.~\ref{M_c1TP.fig}.
Thus in the solar-metallicity case 56 Peg A is well below the luminosity
defining the onset of the TP-AGB.
The situation is similar for the $Z=0.008$, 2.5~\Msun\ case (left panel of
Fig.~\ref{HRtracks.fig}): for a core mass of 0.52~\Msun\ at the onset of the
TP-AGB (see Fig.~\ref{M_c1TP.fig}), Bl\"ocker's core-mass -- luminosity
relationship predits $\log L = 3.3$, larger than the estimated stellar
luminosity of $\log L = 2.7$.
The more complex core-mass -- luminosity relationship of Wagenhuber \& 
Groenewegen \cite{Wagenhuber-1998} yields basically the same results.
Even considering TP-AGB stars in  the luminosity dips following a thermal
pulse (rather than in the quiescent H-burning interpulse stage, as
implicitly assumed by the above core-mass -- luminosity relationships)
would not change the above conclusion that 56~Peg~A cannot be a TP-AGB star,
since these dips are about 0.3~dex fainter in
$\log L$ (see Wagenhuber \& Groenewegen \cite{Wagenhuber-1998}). 

The conclusion that 56~Peg~A cannot be a TP-AGB star plays a central role in the
difficulty of finding for the 56~Peg system a suitable evolutionary channel which is
consistent with the component masses derived from the spectroscopic orbit, as discussed in the
next section. 

\section*{\it\normalsize Difficulty with the masses under the corotation hypothesis}
\label{Sect:masses}

\mbox{}\indent The very small velocity semi-amplitude observed by Griffin
(a mere 1.5 km~s$^{-1}$) yields a mass function of only
3.73~10$^{-5}$~M$_\odot$.
Under the assumption that the giant star's rotation is synchronized with the
orbital motion, and that both motions are in the same plane,
Griffin uses the rotational velocity of 4~km~s$^{-1}$ (as
derived by de Medeiros \& Mayor \cite{deMedeiros-Mayor-1999}) to obtain $\sin i \sim 0.22$
(i.e., $i \sim 13^\circ$). The Hipparcos astrometric data are unfortunately of no help to
corroborate this value of the orbital inclination,
as the sought astrometric orbit
should have $a_1 = 1$~mas, just beyond the Hipparcos detection
capabilities (especially when the orbital period is as short
as 111~d; see Jancart et al. \cite{Jancart-2005}). The masses of the
components are then  constrained by the relation
\begin{equation}
\frac{M_2^3}{\left(M_1+M_2\right)^2}\approx0.0034 \quad\quad {\rm when} \quad\quad i = 13^\circ.
\label{masses.eqn}
\end{equation}
From this relationship, Griffin derives a mass ratio
$q = M_1/M_2$ of the order of 10 (where $M_1$  and $M_2$ are the masses
of the giant component and of its companion, respectively), and lists
pairs $(M_1, M_2) = (2, 0.26), (4, 0.40)$, and (6, 0.52)
M$_{\odot}$ as examples of possible solutions. The author also notes that
these figures rely heavily  on the above assumptions, but does not  feel
the need to reject them as they are indeed required for the system to remain
detached in its present state.

However, potential difficulties with the former evolution of
the system inspired the authors of
the present note to look more deeply into this result from a stellar
evolution standpoint. More specifically, the possible masses derived from
the mass function for the two components (under the assumption of $i =
13^\circ$) raise several questions regarding the past history of the
system, especially when considering the fact that the location of 56~Peg~A in the H-R diagram
forbids it from being a TP-AGB star, as discussed earlier. 
The very barium nature of 56~Peg~A then
implies that it has been polluted through mass transfer from a former TP-AGB companion. The
present  companion star has then to be a carbon-oxygen white dwarf (CO WD). This restricts the
companion's mass range to $M_2
\gtrsim 0.45$~M$_{\odot}$. Therefore, among the three pairs of masses suggested by Griffin, only
the high mass range is in fact relevant to 56~Peg. A semi-empirical
initial--final mass relation for WDs may then be used to further
constrain $M_2$.

Taken together with the most recent initial--final mass relation of
Weidemann \cite{Weidemann-2000}, 
expression (\ref{masses.eqn}) seems to require that the
mass of the giant has been substantially altered by the mass transfer.
Even so, however, there is hardly a way towards an acceptable solution.

\begin{figure}
\resizebox{\hsize}{!}{\includegraphics[angle=-90]{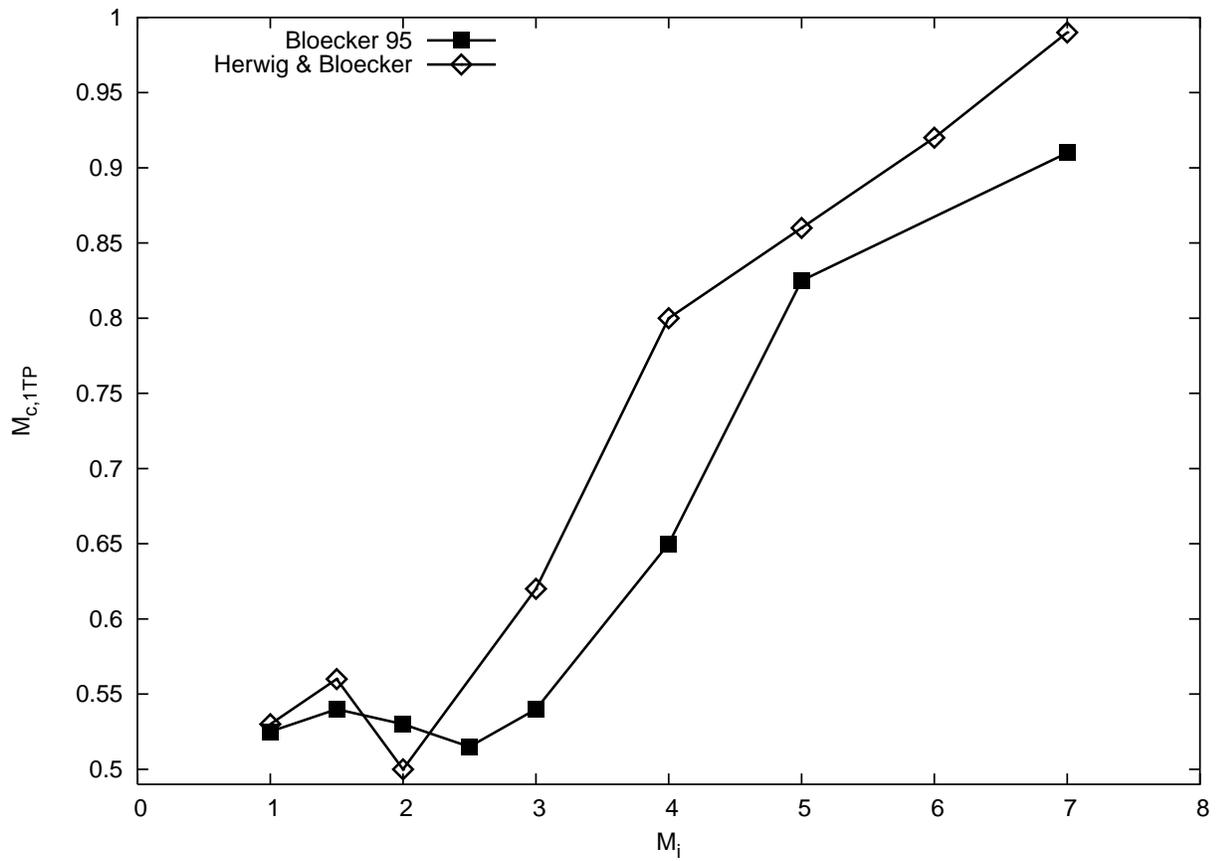}}
\caption[]{
Mass of the CO core at the first thermal pulse, $M_{\rm c,1TP}$, as a
function of initial mass $M_{\rm 2,i}$, according to Bl\"ocker
\protect\cite{Blocker-1995} 
-- line with filled squares -- and to Herwig \& Bl\"ocker (priv. comm. to
Weidemann \protect\cite{Weidemann-2000}) -- line with rhombs. }
\label{M_c1TP.fig}
\end{figure}

As an illustrative example, assume $M_{\rm 2,f} = M_{\rm WD} =
0.68$~M$_{\odot}$. Then from Weidemann relation, $M_{\rm 2,i} =
3$~M$_{\odot}$ and from the mass function, $M_{\rm 1,f} =
8.9$~M$_{\odot}$. The former primary has thus lost in total $3 - 0.68 =
2.32$~M$_{\odot}$. Even assuming that all this mass was dumped on the
companion yields an initial mass of
$M_{\rm 1,i} = 8.9 - 2.32  = 6.58$~M$_{\odot}$ for the progenitor of the
current giant star. But this mass is still larger than 3~M$_{\odot}$, the
estimated initial mass of the former {\em primary}.
This solution is thus not acceptable,
as it is obviously incompatible with the constraint that the WD progenitor
must have been initially the more massive component.
It is in fact possible
to demonstrate graphically that there are no
solutions  for the whole range of relevant masses
(the method is presented in what follows, where it is used with even more
liberal assumptions on the theoretical final masses).
Then we are left with the hypothesis that the initial--final mass
relationship does not apply here.
It is not surprising indeed, that in a mass-transfer system, a 
standard single-star $M_{\rm i}$--$M_{\rm f}$
relation does not hold. We nevertheless note that the
precursor of the current  WD had to pass through the TP-AGB phase, in order to
produce the heavy elements now observed  in the atmosphere of the barium star. This again
constrains the WD mass, as it is required that this mass be no smaller than the CO core mass at
the first thermal pulse on the AGB,
$M_{\rm c,1TP}(M_{\rm 2,i})$. As indicated in the notation, this quantity
is a function of the initial mass of the star. This relationship can be
derived from stellar evolution theory and  
various sources give similar results -- see e.g., Fig.~1 of Weidemann \cite{Weidemann-2000}. 
To fix the ideas, Fig.~(\ref{M_c1TP.fig}) presents  
$M_{\rm c,1TP}$ as a function of $M_{\rm 2,i}$ from Bl\"ocker \cite{Blocker-1995} 
(it is a compilation of his own data with those of Lattanzio \cite{Lattanzio-1986}) and from
calculations by Herwig \& Bl\"ocker (priv. comm. to Weidemann \cite{Weidemann-2000}).
Other theoretical results fall generally in between these two lines, so the range displayed on
Fig.~\ref{M_c1TP.fig} gives an estimate of the theoretical uncertainty.
For our purpose the lowest theoretical predictions of 
$M_{\rm c,1TP}$ for any given $M_{\rm 2,i}$ are relevant, as they are
more liberal -- in the sense that they allow a greater range of possible
final solutions. If we conclude that it is hard to find a solution with
Bl\"ocker's curve, it will be even harder for any other existing relation.
So, in what follows, Bl\"ocker's relation \cite{Blocker-1995} will be used.

\begin{figure}
\resizebox{\hsize}{!}{\includegraphics[angle=-90]{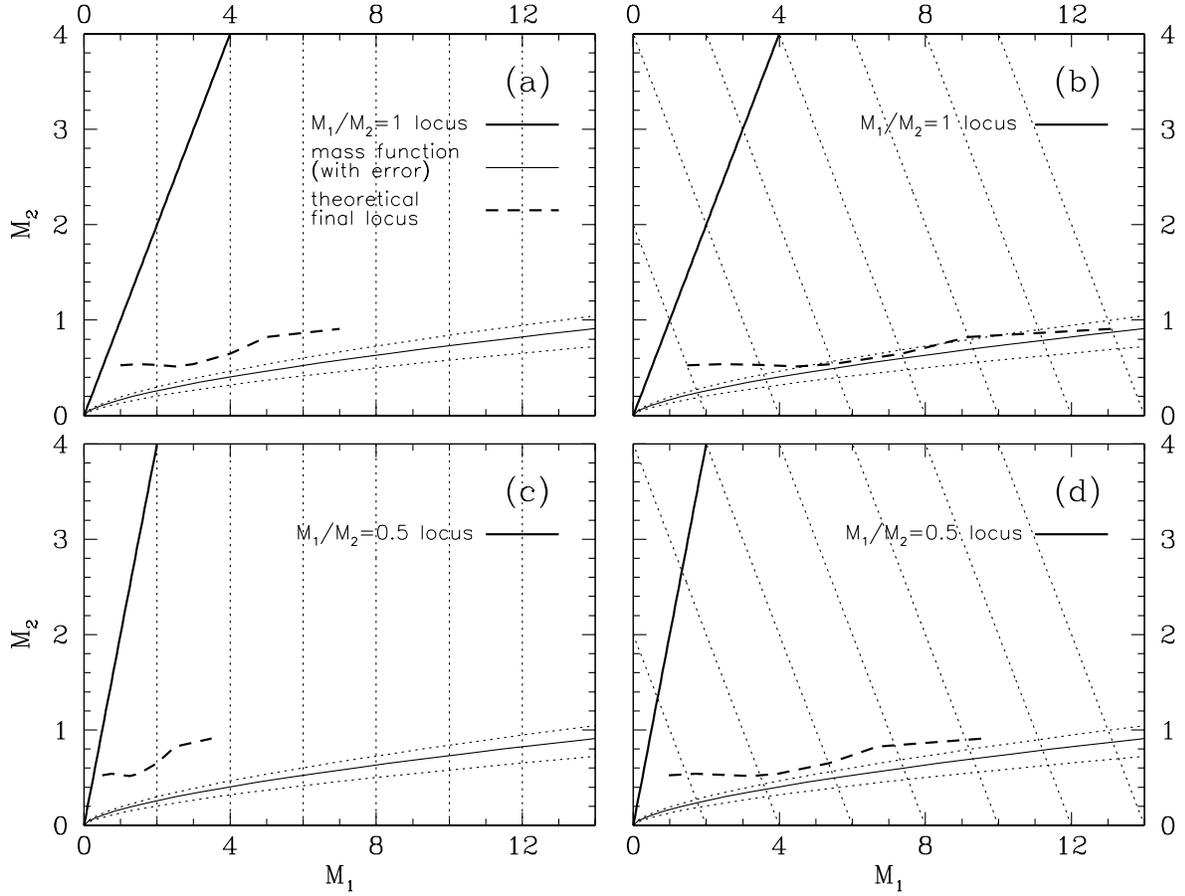}}
\caption[]{
Mass-mass diagrams describing the evolution of the 56~Peg system. The
horizontal and vertical axes correspond to the masses of the current
barium star ($M_1$) and of the current WD ($M_2$), respectively. Each
panel corresponds to a specific evolutionary scenario: (a)
$q_i = M_{\rm 1,i}/M_{\rm 2,i} = 1$, mass loss only (no mass exchange);
(b) $q_i = 1$, conservative mass transfer; (c) $q_i = 0.5$, mass loss only;
(d) $q_i = 0.5$, conservative mass transfer. See text for details. }
\label{extremes.fig}
\end{figure}

As can be seen, the absolute lower limit on the companion's mass $M_2$ is
now raised to $\sim 0.52$~M$_{\odot}$ and this value increases further
with increasing $M_{\rm 2,i}$. This makes the present barium star quite
massive ($M_1 \gtrsim 6$~M$_\odot$ when the nominal inclination derived by
Griffin is adopted). In fact, it becomes impossible to account for
the formation of such a system in the framework of current stellar
evolution theory. This difficulty is illustrated in
Fig.~\ref{extremes.fig}, presenting mass-mass diagrams for the
components of the 56~Peg system, for four evolutionary scenarios. 
In this figure,
the horizontal axis corresponds to the mass  $M_1$ of the now-barium star,
and the vertical axis to the mass  $M_2$ of the now-WD.
The actual evolution of the system would be represented
by a curve joining points corresponding to initial and final configurations.
As the exact course of the 56~Peg evolution is unknown, we consider two
extreme scenarios, which should encompass the actual evolution: 
no mass transfer, i.e., only mass loss from
$M_2$ (panels a and c) or conservative mass transfer from $M_2$ to $M_1$,
i.e., no systemic mass loss (panels b and d).
With the assumed mass-loss/mass-transfer modes, the system evolves along
straight lines  in the ($M_1, M_2)$ diagram: vertical for pure
mass loss, or inclined at
$-45^\circ$ for conservative mass transfer. To guide the eye, the course
of evolution is exemplified by several thin dotted lines in each panel.

Adopting a given value for the {\em initial} mass ratio $q_i =
M_{\rm 1,i}/M_{\rm 2,i}$, one may then represent the initial states of the
system for any given
$q_i$ as straight lines of appropriate inclination going through (0, 0).
These are shown as thick lines, for $q_i = 1$ in panels a and b, and
for $q_i = 0.5$ in panels c and d. Note that $q_i = 1$ is the limiting
case, as initially
$M_2$ had to be at least slightly larger than $M_1$ in order for the
primary component to evolve faster.

Two independent constraints -- which should be consistent with each other
-- are available  to characterize the {\em final} (i.e., present) state 
of the binary system. First, the mass function 
expressed by Eq.~\ref{masses.eqn} corresponds to the masses consistent
with the observed radial-velocity curve and the adopted orbital
inclination. This relationship between
$M_{\rm 1,f}$ and $M_{\rm 2,f}$ is shown in each panel as a thin solid
line, accompanied by thin dotted lines representing the 1$\sigma$
errors (which combine the uncertainties on the mass function and on the
inclination). Second, the theoretical relation $M_{\rm c,1TP}(M_{\rm
2,i})$ of Bl\"ocker (Fig.~\ref{M_c1TP.fig}) provides another estimate
of $M_{\rm 2,f}$ (thick broken line) that connects to the initial value
$M_{\rm 2,i}$ through the thin dotted lines describing the
mass-loss/mass-transfer evolution. Only points at or above this line
represent possible final configurations. 
Note that the transformation of this theoretical final locus due to the
changes in assumptions is a simple stretching and squeezing along
the horizontal axis.

From Fig.~\ref{extremes.fig}, it is obvious that a marginal agreement
between the observed and theoretical final masses is possible only in the
most extreme case ($q_i=1$, conservative mass transfer: panel~b of
Fig.~\ref{extremes.fig}). But conservative mass transfer originating from
the convective envelope of a TP-AGB star (the progenitor of the current
WD) does not seem very plausible, as such a mass-transfer mode would
rapidly suffer from dynamical instabilities. Another difficulty with this
solution is the rather large mass implied for the
current giant star ($M_1 \ge 5$~M$_{\odot}$). 
According to the stellar-evolution grid of Lejeune \& Schaerer \cite{Lejeune-Schaerer-2001} 
displayed in
Fig.~\ref{HRtracks.fig}, such massive giant stars (of solar or subsolar
metallicity) have absolute magnitudes well in excess of the value 
inferred for 56~Peg~A.

In fact, these tracks impose $M_1 \le 4.5$~M$_{\odot}$ for 56~Peg~A.
Such a limit on $M_1$ can be inferred by inspection of the
local $M_V$ minima at the base of the giant branch:  tracks for any star
more massive than 4.5~\Msun\ will never reach as low a luminosity as
$M_V \sim -1.3$ on the giant branch, irrespective of the metallicity.
Thus the high mass solutions inferred from the spectroscopic
orbit may be excluded right away from luminosity considerations.
 
Considering the error bars on 56~Peg photometry and metallicity, the mass range 2 -- 4 \Msun\
appears likely for the giant star (Fig.~\ref{HRtracks.fig}).   
Then from Eq.~\ref{masses.eqn} (corresponding to an orbital inclination of $i = 13^\circ$), it
follows that $M_2$ lies in the range 0.26--0.40 \Msun\ -- much too small for a CO WD.

\section*{\it\normalsize Towards a new model}
\label{Sect:newmodel}

\mbox{}\indent How is it then possible to reconcile theory and observations? 
The $V \sin i$ estimate of de Medeiros \& Mayor \cite{deMedeiros-Mayor-1999} seems beyond doubt,
even though it involves some assumptions on the macroturbulent
contribution to the observed line width. An independent
determination by Gray \cite{Gray-1989}, using the Fourier transform technique which
makes it possible to separate the macroturbulent from the rotational
contribution, yields 3.9~\kms, very close to the 4.4~\kms\ value of de
Medeiros \& Mayor \cite{deMedeiros-Mayor-1999}.

One simple possibility would then be to allow $\sin i$ outside the  $1
\sigma$ range. In this case lower $\sin i$ values are needed: the more
edge-on the system is, the higher the companion mass becomes and the
higher   the observational ($M_{\rm 1,f}, M_{\rm 2,f}$) locus lies on
the mass-mass diagram.

Dropping the assumption of corotation,
involved in deriving the value of $\sin i$, appears to offer
very promising prospects, as we now show. This would improve the
situation only  if the giant rotates {\em faster} than corotation, 
implying a larger equatorial velocity. The inclination derived from
the same value of the projected rotational velocity as adopted by Griffin
would then become {\em smaller}. The observational final locus of 56~Peg
on the $(M_1, M_2)$ diagram would then come closer to our theoretical
limit. To fix the ideas, for the
$q_i = 0.5$ and pure mass-loss case (case c),  consistency between the
mass-function and $M_{\rm c,1TP}(M_{\rm 2,i})$ loci can be achieved if $i
= 5^\circ$. This in turn implies an equatorial rotation velocity of about
48 \kms! The only way to reach such a fast rotation is by spin accretion
during the mass-transfer episode \cite{Theuns-1996,Jeffries-Stevens-1996}.
The spin accretion had to occur recently, because
the dynamo associated with such a fast
rotation and the resulting magnetic braking would slow the
rotation down on a time scale of the order of 10$^8$~yr (see the paper by
Theuns et al. \cite{Theuns-1996} and
references therein). One similar case is known among barium stars, namely
HD~165141, a rapidly-rotating ($V \sin i = 14$~\kms) star, also
classified \cite{Theuns-1996,Fekel-1993,Jorissen-1996} as RS~CVn
despite its long orbital period of 5200~d. The WD companion is
the warmest known ($T_{\rm eff} \sim 35000$~K) among barium-star
companions, and its rather short cooling time scale (about 10$^7$~y)
clearly hints at a recent mass (and spin) accretion event.
One notes a large difference in orbital period between this system and
56~Peg, which arguably somewhat limits the resemblance. Spin accretion
in HD~165141 is thought to occur through wind and accretion disc
formation while the separation of components in 56~Peg is comparable
to (or even smaller than) a typical AGB giant radius. Thus a mass transfer
mode other than wind accretion had to be involved in the 56~Peg system.
But it is worth noting that fast rotation (up to 100~\kms!) is in fact
common among at least one class of related systems, namely d'-type
symbiotics \cite{Jorissen-2005}. These objects also do exhibit
some degree of barium enrichment. The only d'-type symbiotic with
known orbital period, V417 Cen, has $V \sin i = 70$~\kms~and
$P=247^{\rm d}$, which is much closer to the orbital period value
of 56~Peg. That proves that spin accretion is somehow possible (and
efficient!) even in systems with such relatively short orbital periods.

So, could 56~Peg be a similar case? Very likely, as 
\begin{itemize}
\item[(i)] the strong UV continuum \cite{Schindler-1982a,Schindler-1982b} 
has been ascribed to a warm WD of $T_{\rm eff} = 32000 \pm
4000$~K and $M_2 = 1$~\Msun.
Sweeney \cite{Sweeney-1976} predicts a cooling
time of less than $50\;10^6$~y for a WD of  that temperature. This cooling time is shorter than
the magnetic braking time scale of the giant, so that the latter is still rotating fast.
\item[(ii)] many different observers
have reported signatures of chromospheric activity (see Griffin
\cite{Griffin-2006} and references therein).
\end{itemize}

\begin{figure}
\resizebox{\hsize}{!}{\includegraphics{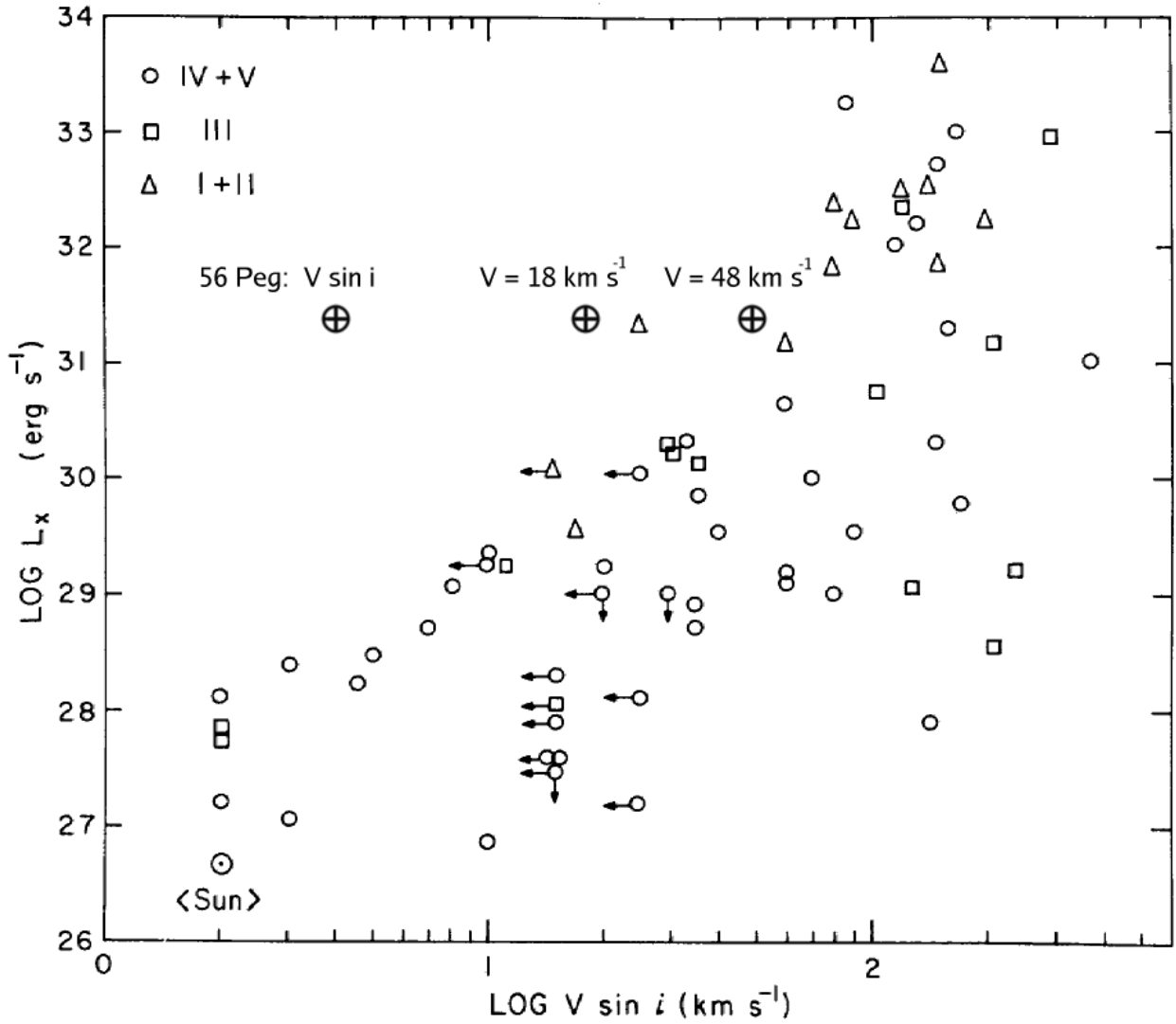}}
\caption[]{X-ray luminosities vs.\ projected rotational velocities for
56 Peg (encircled crosses, for $V \sin i$ and for two values of $V$:
assuming orbital corotation or fast rotation) and stars of various spectral
types and luminosity classes
detected by the {\it Einstein} Observatory. Different symbols indicate
different luminosity classes.  Adapted from Pallavicini
et al.~\protect\cite{Pallavicini-1981}. }
\label{Fig:Pallavicini}
\end{figure}

To mention just a few, Schindler et al. \cite{Schindler-1982a},
quoting Linsky et al. \cite{Linsky-1979}, indicate that `56 Peg exhibits an unusual
Ca~II K~line emission profile for its MK spectral type (...) The Ca II
emission more nearly resembles that seen in solar plages.' Eaton \cite{Eaton-1995}
has remarked that the H$\alpha$ line profile is much shallower than the
same line in inactive stars, thus suggesting that H$\alpha$ has been
filled in by emission. Similar indications of emission exist for the
ultraviolet  Mg~II h \& k~doublet and the Na~I D~line (see Griffin
\cite{Griffin-2006} and
references therein). The question nevertheless arises whether this
activity is triggered by fast rotation or by mass transfer
in the binary system. Several arguments support the hypothesis of
a fast rotator.  The first is the constancy of the Ca~II H \& K emission
profiles over several years \cite{Schindler-1982a}, whereas a mass-transfer origin for these
lines would rather cause some variability, associated with irregularities in the mass
transfer.\footnote{Surprisingly, the Mg~II  h \& k~doublet, which is supposed to form at somewhat
higher chromospheric levels than the CaII H \& K, according to Mullan \& 
Stencel \cite{Mullan-Stencel-1982}, is strongly variable
\cite{Schindler-1982a,Mullan-Stencel-1982,Stencel-1984}. This variability cannot be
ascribed to orbital modulation, since the system is seen nearly pole-on in the present model.
Variations in the outflowing wind of the K giant, where the  Mg~II doublet supposedly
forms, represent therefore an interesting possibility, as advocated by Mullan \& 
Stencel \cite{Mullan-Stencel-1982}
and Schindler et al. \cite{Schindler-1982a}.} 
Second, the bright UV emission lines of C~II, C~III, C~IV, He~II,
Si~II, Si~III, Si~IV, N~III and N~V observed by Schindler et al. \cite{Schindler-1982a} 
along with the X-ray
flux are typical signatures of a hot corona associated with magnetic heating
and fast rotation. Schindler et al. \cite{Schindler-1982a} felt obliged to
reject this obvious explanation, since the observed $V \sin i$ is low.
But the present paper precisely argues that these difficulties, as well as
those raised by Griffin's orbital elements, vanish if one assumes that the
star is a fast rotator. Very interestingly, Schindler et al.
\cite{Schindler-1982b} argue that 56~Peg deviates from the rotation-activity
($L_X$ versus $V \sin i$) correlation of Pallavicini et al.
\cite{Pallavicini-1981}.
Now it becomes obvious that it is the small $\sin i$ value that drags
this object that far to the left of the graph.
Using corotation velocity (i.e.\ $V = $ 18~\kms) instead of $V \sin i$
improves the agreement, but 56 Peg is still an outlier (and would be even
more so if one could resolve $V$ from $\sin i$ for the other objects).
However, adopting $V \sim $ 50~\kms\ would restore 56~Peg right within the
expected locus of RS~CVn binaries, as far as its X-ray flux is concerned!
This is shown in Fig.~\ref{Fig:Pallavicini}, which compares the X-ray flux
of 56~Peg (corrected according to the new distance estimate from Hipparcos:
165~pc instead of 215~pc as estimated by Schindler et al.
\cite{Schindler-1982a}) to that of late-type (including active) stars. 

Given the small orbital separation, one may still wonder whether there 
should not be as well some contribution to the X-ray flux coming 
from accretion, as advocated by Schindler et al. \cite{Schindler-1982a}
and Dominy \& Lambert \cite{Dominy-Lambert-1983}. 
The major evidence suggesting that there may be a link between mass 
transfer and X-ray emission in active binaries is the correlation 
between the X-ray luminosity and the Roche-lobe filling fraction 
$\Gamma_2$ for 
RS~CVn systems noted by Welty \& Ramsey \cite{Welty-Ramsey-1995}. 
Singh et al. \cite{Singh-1996} re-examined this issue using samples of RS~CVn 
and Algol binaries, and confirm the weak 
correlation found earlier by Welty \& Ramsey \cite{Welty-Ramsey-1995}. 
However, Singh et al. ``{\it regard this as a rather 
weak argument, because both the X-ray luminosities and the Roche lobe 
filling fractions are themselves correlated with the stellar radii in 
the sample of RS~CVn binaries, and thus regard this correlation as 
merely a by-product of the inherent size dependence of these 
quantities.''} 

Actually, despite the short orbital period, the filling factor of
56~Peg~A is not that close to
unity,\footnote{Adopting $i = 5^\circ$, $M_1 = 3$~\Msun, $M_2 = 0.96$~\Msun,
Griffin's orbital
elements yield an orbital separation $A = 152$~R$_{\rm \odot}$ and a Roche
radius around the giant component of 73~\Rsun, well in excess of the
40~\Rsun\ representing the stellar radius itself, and translating into
$\Gamma_2 = 0.55$.
One may note, however, that rotation faster than orbital motion complicates
somewhat the picture: the usual Roche geometry does not strictly apply,
hence the effective $\Gamma_2$ will be a bit larger \cite{Kruszewski-1963}.}
and therefore mass transfer to the companion occurs only through the
accretion of the giant wind, which is quite moderate according
to Reimers' mass-loss law: a mere
$2.7\;10^{-9}$~\Msun~y$^{-1}$ (adopting for the giant $M_1 = 3$~\Msun,
$R_1 = 40$~R$_{\odot}$ and 
$\log L/L_{\odot} = 2.7$). Schindler et al. \cite{Schindler-1982a} and Dominy \& 
Lambert \cite{Dominy-Lambert-1983} showed that
such a mass loss rate could indeed account for the observed X-ray flux in a 100~d system, using
the Hoyle \&  Lyttleton \cite{Hoyle-Lyttleton-1939} 
accretion cross section. They failed, however, to recognize that
this accretion cross section overestimates by about one order of magnitude the actual cross
section, as inferred from hydrodynamical 
simulations \cite{Theuns-1996,Mastrodemos-Morris-1998,Nagae-2004}.   
This constitutes a further argument against a strong contribution
from mass accretion to the X-ray flux.

Symbiotic systems provide another interesting sample to which 56~Peg
may be compared. One of the defining properties of symbiotic
systems is to host a hot compact star, mostly a WD,
accreting matter from a giant companion \cite{Corradi-2003}. The 
WD is heated either directly by accretion, or indirectly 
by nuclear burning fueled by accretion (see Jorissen et al. \cite{Jorissen-2003a} 
and references therein). 
Symbiotic stars are X-ray sources \cite{Murset-1997}, at the level $10^{30}$ to 
$10^{33}$~erg~s$^{-1}$, to be compared with 
$2.3\;10^{31}$~erg~s$^{-1}$ for 56~Peg \cite{Schindler-1982a,Voges-1999,Schwope-2000} 
(corrected for the new Hipparcos distance).  The physical process emitting these X-rays in
symbiotic stars is still  debated. Direct evidence for accretion disks, 
in the form of continuum flickering and far UV continuum 
(and, by extension, X-rays) is not 
usually found in symbiotic stars \cite{Sokoloski-2001,Sokoloski-2003,Sion-2003},
unlike the situation prevailing in cataclysmic variables and low-mass 
X-ray binaries \cite{Sokoloski-2001,Guerrero-2001}. 
Other mechanisms were therefore advocated to account for the X-ray 
emission from symbiotics \cite{Murset-1997}, like 
thermal radiation from the hot component in 
the case of supersoft X-ray sources, or the shock forming in 
the collision region between the winds from the hot and cool 
components. Interestingly, Soker \cite{Soker-2002} even suggests that 
the X-ray flux from symbiotic stars exclusively arises from the fast 
rotation of the cool component spun up by wind accretion from the 
former AGB component (now a WD)! If that hypothesis is correct, the 
properties of X-rays from symbiotics would be undistinguishable from 
those of active binaries. 

To conclude, it is thus unlikely that mass transfer possibly 
occurring in the 56 Peg system contributes significantly to the 
observed X-ray flux. The X-rays from 56~Peg share many properties of 
active binaries. On top of  the arguments mentioned above, the ROSAT $HR1$
hardness ratio  for 56~Peg, as given by Voges et al. \cite{Voges-1999} 
($HR1 = 0.68$, where $HR1 = (H-S)/(H+S)$ and $H$ and $S$ are the ROSAT PSPC
count rates in the 0.5 -- 2.0~keV and 0.1 -- 0.4~keV ranges, respectively)
does lie in the range seen in RS~CVn stars taken from the list of
Singh et al. \cite{Singh-1996}.

Another argument supporting the presence of rotation-driven activity in 56~Peg 
is provided by the surface flux $S_{\rm C IV}$ of the C~IV~$\lambda 155$~nm 
line, which falls exactly on the relation $S_{\rm C IV} - P_{\rm rot}$ derived 
by Gunn et al. \cite{Gunn-1998} for active binary stars. Adopting for
56~Peg~A a radius of 40~\Rsun\ and a distance of 165~pc, 
the observed \cite{Schindler-1982a} C~IV flux 
[$1.2\;10^{-12}$~erg~cm$^{-2}$~s$^{-1}$]
translates into a surface flux of $4.0\;10^{4}$~erg~cm$^{-2}$~s$^{-1}$,
as compared to $5.1\;10^{4}$~erg~cm$^{-2}$~s$^{-1}$ from the period
-- surface-flux relationship of Gunn et al. \cite{Gunn-1998} for an estimated rotation period
of $P_{\rm rot} = 42$~d, adopting $V_{\rm rot} = 48$~\kms. Similarly, the
$S_{\rm X}/S_{\rm bol}$ and $S_{\rm C IV}/S_{\rm bol}$ ratios for
56~Peg fall on the tight relation obtained by Dempsey et al.
\cite{Dempsey-1993} for RS~CVn systems.

\section*{\it\normalsize Conclusions}

\mbox{}\indent The following coherent picture thus emerges  for the 56~Peg
system:
\begin{itemize}
\item Evolutionary tracks combined with Tycho-2 photometry  and
Hipparcos parallax ($<V> = 4.76$, $<B-V> = 1.28$, $\varpi = 6.07\pm0.67$~mas,
$M_V = -1.32{+0.23 \atop-0.26}$) imply that the giant component in the 56~Peg
system has a mass of the order of 2 to 4
\Msun, for metallicities solar or slightly subsolar ($-0.4 \le [{\rm Fe/H}]
\le 0.$). At solar metallicity, the Lejeune \& Schaerer
\cite{Lejeune-Schaerer-2001} tracks predict $M = 3$~\Msun for our adopted
56~Peg~A values
of $T_{\rm eff} = 4200$~K and $\log L = 2.7$, corresponding to
$M_{\rm bol} = -2.0$.
\item Whatever its metallicity in the above range, 56~Peg~A is not luminous
enough to be on the TP-AGB. The barium overabundance observed in the atmosphere
of 56~Peg~A must therefore be ascribed to mass transfer in the binary system.
\item The WD companion of temperature 32000~K (inferred from a fit to the
observed UV continuum) implies that the WD is young (with a cooling time
scale of the order of $50\;10^6$~y).
56~Peg~A might therefore have been spun up during the mass transfer process
(as d'-type symbiotics of similar orbital periods, thus representing yet
another example of the class of WIRRING systems defined by Jeffries \&
Stevens \cite{Jeffries-Stevens-1996}) and not yet slowed down by magnetic
braking, which occurs on a time scale longer than the WD cooling time scale.
In this scenario, high rotational velocity is achieved after and
independently of tidal orbital circularization.
\item 56~Peg~A is therefore an active star, as
evidenced  by many emission lines in the optical (H$\alpha$, Na~D, CaII H \& K) and ultraviolet.
Diagnostics like $S_{\rm X}/S_{\rm bol}$, $S_{\rm C IV}/S_{\rm bol}$ (where $S$ denotes the
surface flux) and $L_X$ versus $V_{\rm rot}$ all fall along the relations expected for active
stars of the RS~CVn type.
The X-ray flux is therefore attributed to the activity of the giant star rather than to mass
transfer in the binary system.
\item The strong variability of the Mg II h \& k lines has been ascribed to variability in the
wind from the K giant. The above stellar parameters predict $3\;10^{-9}$~\Msun~y$^{-1}$ from the
Reimers mass-loss law.
\item All activity indicators are consistent with a rotational velocity of
30--50~km~s$^{-1}$, implying an orbital inclination $i \sim$
5--8$^\circ$.
\item The orbital mass function combined with the giant mass range (2--4~\Msun) then implies 
$M_2 = 0.75$ to 1.15~\Msun, now consistent with it being a CO WD.
\item If the system is seen nearly pole-on, Griffin's orbital elements then yield an orbital 
separation $A = 152$~R$_\odot$ (for $M_1 = 3$~\Msun\ and $M_2 = 0.96$~\Msun), implying a Roche
radius around the giant component of 73~\Rsun, well in excess of the 40~\Rsun\ representing the
stellar radius itself. The short orbital period (111~d) is not at all incompatible
with the system being currently detached.
\item The assumed inclination, $i = 5^\circ$, has an {\it a priori}
probability of one part in 260. Admittedly, it is a small value, but it does
not make the proposed solution impossible. In the light of the presented
evidence, and bearing in mind the opening quotation from Sir Arthur Conan
Doyle, the authors feel encouraged enough to submit their reasoning under
the reader's judgement.

\end{itemize}

{\it Acknowledgements}\\

\mbox{}\indent
We thank Dr. R. Griffin for his patience which allowed this note to be
published in the same issue of this {\it Magazine} as his orbital elements.
A.J. is Senior Research Associate from F.N.R.S. (Belgium) and
A.F. is Foreign Postdoctoral Fellow from F.N.R.S. (Belgium) under grant
1.5.108.05 F.

\bibliographystyle{unsrt}

\end{document}